\documentclass[aps,prl,twocolumn,showpacs]{revtex4-1}
\usepackage{amsmath}
\usepackage[dvips,xdvi]{graphicx}
\usepackage{amssymb}
\usepackage{epsfig}
\begin{document}
\title{Quantum quench and non-equilibrium dynamics in lattice-confined spinor condensates}
\author{Z. Chen}
\author{T. Tang}
\author{J. Austin}
\author{Z. Shaw}
\author{L. Zhao}
\author{Y. Liu}
\email{Electronic address: yingmei.liu@okstate.edu}
\affiliation{Department of Physics, Oklahoma State University,
Stillwater, Oklahoma 74078, USA}
\date{\today}
\begin{abstract}
We present an experimental study on non-equilibrium dynamics of a spinor condensate after it is quenched across a
superfluid to Mott insulator (MI) phase transition in cubic lattices. Intricate dynamics consisting of spin-mixing
oscillations at multiple frequencies are observed in time evolutions of the spinor condensate localized in deep lattices
after the quantum quench. Similar spin dynamics also appear after spinor gases in the MI phase are suddenly moved away
from their ground states via quenching magnetic fields. We confirm these observed spectra of spin-mixing dynamics can be
utilized to reveal atom number distributions of an inhomogeneous system, and to study transitions from two-body to
many-body dynamics. Our data also imply the non-equilibrium dynamics depend weakly on the quench speed but strongly on the
lattice potential. This enables precise measurements of the spin-dependent interaction, a key parameter determining the
spinor physics.
\end{abstract}

\maketitle

A spinor Bose-Einstein condensate (BEC) is a multi-component condensate possessing a spin degree of
freedom~\cite{StamperKurnRMP}. Combined with optical lattices and microwave dressing fields, spinor gases offer an
unprecedented degree of control over many parameters and have thus been considered as ideal candidates for studying
complicated non-equilibrium
dynamics~\cite{StamperKurnRMP,spinorquench,quench,ZhaoUwave,Zhao2dLattice,LatticeSpinor,Greiner01,Triangular,Hexagon,KSpinDynamics,CrSpinDynamics,JiangGS}.
Such a spinor system can be easily prepared far away from equilibrium through quenching one of its highly-controllable
parameters, e.g., the number of atoms, temperature, total spin of the system, the lattice potential, or the dimensionality
of the
system~\cite{StamperKurnRMP,spinorquench,quench,ZhaoUwave,Zhao2dLattice,LatticeSpinor,Triangular,KSpinDynamics,CrSpinDynamics,JiangGS}.
Interesting dynamics have also been initiated in lattice-confined spinor gases by non-equilibrium initial states, such as
interaction-driven revival dynamics in one-dimensional Ising spin chains~\cite{IBloch02}, dynamics and equilibration of
spinor BECs in two-dimensional lattices~\cite{Zhao2dLattice}, and spin-mixing dynamics of tightly confined atom pairs in
cubic lattices~\cite{U2Deeplattice,U2Deeplattice2}. Another notable advantage of spinor systems on investigating
non-equilibrium dynamics is their long equilibration time, ranging from tens of milliseconds to several
seconds~\cite{StamperKurnRMP,Zhao2dLattice}. Experimental studies on non-equilibrium dynamics have been conducted in
spinor gases extensively at two extremes, i.e., in a clean two-body system with a pair of atoms in the Mott-insulator (MI)
phase~\cite{U2Deeplattice,U2Deeplattice2}, and in a many-body system with more than $10^4$ atoms in the superfluid (SF)
phase~\cite{StamperKurnRMP,ZhaoUwave,Zhao2dLattice,Triangular}. Transitions between these two extremes, however, remain
less explored~\cite{spinorquench}.

In this paper, we experimentally confirm that lattice-trapped spinor BECs provide a perfect platform to understand these
less-explored transitions. Our experiments are performed in a quantum quench scenario starting with an antiferromagnetic
spinor BEC at its SF ground state, based on a theoretical proposal in Ref.~\cite{spinorquench}. We continuously quench up
the potential of a cubic lattice to a very large value, which completely suppresses tunnellings to freeze out atom number
distributions in individual lattice sites. Rich spin dynamics are observed at fast quench speeds and adiabatic SF-MI
quantum phase transitions are detected after sufficiently slow lattice ramps. About half of the data shown in this paper
are collected after the lattice is quenched at an intermediate speed, which is slow enough to prevent excitations to
higher vibrational bands while remaining fast enough to suppress hopping among lattice sites. We observe intricate
dynamics consisting of spin-mixing oscillations at multiple frequencies in spinor BECs after the quantum quench in
magnetic fields of strength $B<60~\mu$T. The rest of our data are taken after adiabatic lattice ramps. Similar spin
dynamics also occur after we abruptly move spinor gases in the MI phase away from their ground states via quenching
magnetic fields. In our system, an inhomogeneous system with an adjustable peak occupation number per lattice site
($n_{\rm peak}$), a significant amount of lattice sites are occupied by more than two atoms. The observed spin-mixing
spectra are thus utilized to study transitions between two-body and many-body spin dynamics and to reveal atom number
distributions of an inhomogeneous system. Our data also indicate the non-equilibrium dynamics depend weakly on the quench
speed but strongly on the lattice potential. Every observed spin dynamics is found to be well described by a sum of
multiple Rabi-type spin-mixing oscillations. This enables us to precisely measure the ratio of the spin-independent
interaction $ U_0 $ to the spin-dependent interaction $U_2$, an important factor determining the spinor physics.

The site-independent Bose-Hubbard model has successfully described lattice-confined spinor
BECs~\cite{spinorquench,JiangLattice,ZhaoSinglet}. We can thus understand our data taken in deep lattices with a
simplified Bose-Hubbard model by ignoring the tunnelling energy $J$ as follows~\cite{spinorquench,ZhaoSinglet},
\begin{equation}
H = \dfrac{U_{0}}{2} n (n-1) + \dfrac{U_{2}}{2} (\vec{S}^{2} -2 n)+ q (n_{1}+n_{-1}) -\mu n~. \label{MF}
\end{equation}
Here $U_{2}$ is positive (negative) in antiferromagnetic (ferromagnetic) spinor BECs, $q$ is the net quadratic Zeeman
energy induced by magnetic and microwave fields, $\mu$ is the chemical potential, $n =\sum_{m_{F}}n_{m_F}$ is the total
atom number in each lattice site with $n_{m_{F}}$ atoms staying in the hyperfine $m_{F}$ state, and $\vec{S}$ is the spin
operator~\cite{spinorquench,ZhaoSinglet}.

\begin{figure}[t]
\includegraphics[width=82mm]{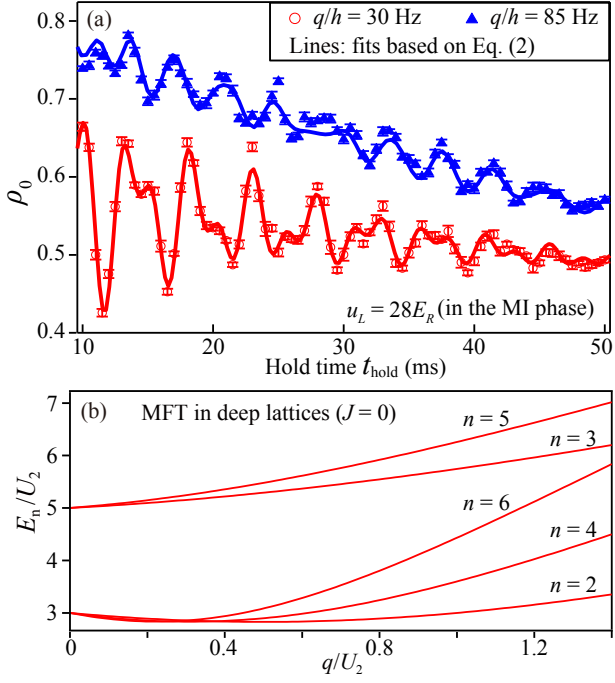}
\caption{(a) Observed dynamics of spin-0 atoms after Quench-Q sequences to different $q$. Lines are fits based on
Eq.~\eqref{fitting}. (b) Lines denote the predicted energy $E_n=h\cdot f_n$ (see text).} \label{quenchQ}
\end{figure}

\begin{figure}[t]
\includegraphics[width=82mm]{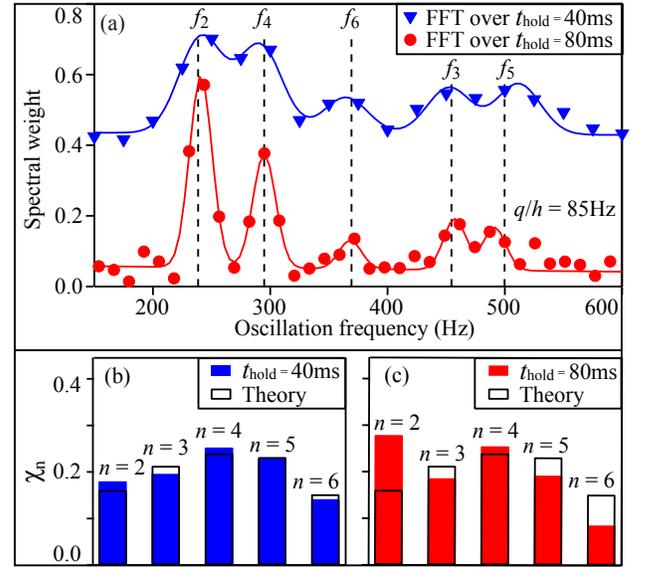}
\caption{(a) Triangles (circles) represent fast Fourier transformations (FFT) over the first 40~ms (80~ms) of $t_{\rm
hold}$ on the $q/h=85$~Hz data set shown in Fig.~\ref{quenchQ}~(a). Vertical lines mark the predicted $f_n$ (see text).
Solid lines are five-Gaussian fits. Results obtained at $t_{\rm hold}=40$~ms are shifted up by 0.4 for visual clarity. (b)
Atom number distributions extracted from the $t_{\rm hold}=40$~ms FFT spectrum in Panel~(a). We define $\chi_n$ as the
fraction of atoms localized in lattice sites having $n$ atoms, and extract $\chi_n$ from dividing the area below the
corresponding peak in a FFT spectrum by the spin oscillation amplitude $D_n$ (see Ref.~\cite{FFT}). Black bars mark the
predicted $\chi_n$ in Mott-insulator shells at $n_{\rm peak}=6$ based on Eq.~\eqref{MF} and the Thomas-Fermi
approximation. (c) Similar to Panel~(b) but extracted from the $t_{\rm hold}=80$~ms FFT spectrum in Panel~(a).}
\label{FFT}
\end{figure}

In our experiment, we start each cycle at $q/h=$40~Hz in free space with a spin-1 antiferromagnetic spinor BEC of up to
$10^5$ sodium ($^{23}$Na) atoms in its ground state, the longitudinal polar (LP) state with $\rho_0=1$ and
$m=0$~\cite{procedure}. Here $\rho_{m_F}$ is the fractional population of the $ m_F$ state, $m=\rho_{+1}-\rho_{-1}$ is the
magnetization, and $ h $ is the Plank constant. Two different quench sequences, Quench-L and Quench-Q, are applied in this
paper. In the Quench-L sequences, we tune magnetic fields to a desired $q$ and then quench up the depth $u_L$ of a cubic
lattice from 0 to $28E_R$ within a time duration $t_{\rm ramp}$, where $ E_R $ is the recoil energy~\cite{procedure}. This
final depth $u_L$ is much larger than SF-MI transition points and thus deep enough to localize atoms into individual
lattice sites. We carefully set $t_{\rm ramp}$ based on two criteria (see Ref.~\cite{ramp}). In the Quench-Q sequences, we
adiabatically ramp up cubic lattices to a final depth of $u_L\geq 28E_R$ in a high field (where $q\gg U_2$), which ensures
atoms cross SF-MI transitions and enter into their ground states (where $\rho_0\simeq 1$) in the MI
phase~\cite{JiangLattice}; and we then suddenly quench magnetic fields to a desired $q$ for initiating non-equilibrium
dynamics. After each quench sequence, atoms are held in lattices for a certain time $t_{\rm hold}$ followed by being
abruptly released from the lattices and detected via the microwave imaging.

\begin{figure*}[t]
\includegraphics[width=178mm]{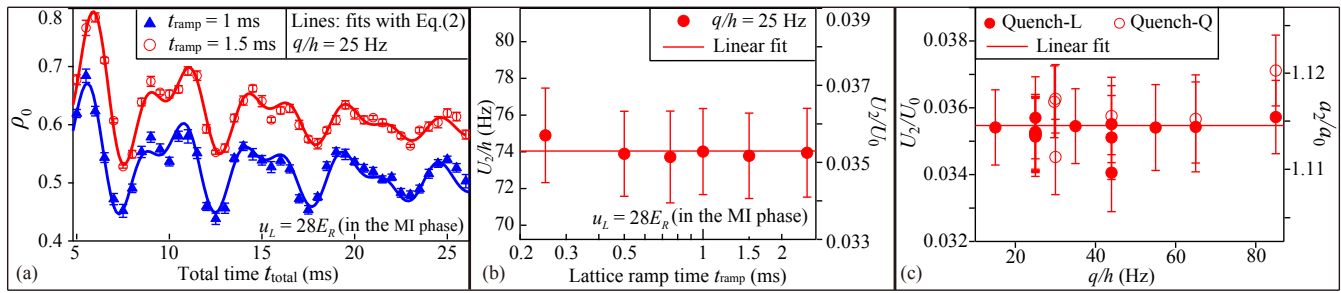}
\caption{(a) Observed spin dynamics after Quench-L sequences at two $t_{\rm ramp}$. Lines are fits based on
Eq.~\eqref{fitting}. Data taken at $t_{\rm ramp}=$~1.5~ms are shifted up by 0.1 for visual clarity. (b) Extracted $U_2$
and $U_2/U_0$ from fitting observed dynamics with Eq.~\eqref{fitting} at various $t_{\rm ramp}$. The horizontal line is a
linear fit. (c) Similar to Panel~(b) but extracted from our data taken under 20 different conditions. The right axis marks
the corresponding ratio $a_2/a_0=(U_2+U_0)/(U_0-2U_2)$, where $a_0$ and $a_2$ are scattering lengths.} \label{U2}
\end{figure*}

Interesting non-equilibrium dynamics consisting of spin-mixing oscillations at multiple frequencies are observed after
both Quench-L and Quench-Q sequences in spinor gases localized in deep lattices at $q/h< 100$~Hz. Two typical time
evolutions detected after Quench-Q sequences are shown in Fig.~\ref{quenchQ}(a). Such an evolution appears to be fit by a
composition of multiple Rabi-type oscillations (see solid lines in Fig.~\ref{quenchQ}(a) and Eq.~\eqref{fitting}). This
can be explained by considering that $n$ atoms tightly confined in one lattice site display a Rabi-type oscillation at a
fixed frequency $f_n$, and the observed dynamics combine all time evolutions occurring in individual lattice sites for our
inhomogeneous system. We derive $f_n=E_{n}/h$ from the mean-field theory (MFT), where $E_{n}$ is the energy gap between
the ground state and the first excited state at a given $n$ (see Fig.~\ref{quenchQ}(b)). Analytical expressions for $f_n$
can be found at $n=2$ and $n=3$, i.e., $f_{2}=U_{2}\sqrt{9-4(q/U_{2})+4 (q/U_{2})^2}/h$ and
$f_{3}=U_{2}\sqrt{25+4(q/U_{2})+4(q/U_{2})^2}/h$. We develop the following empirical formula based on the predicted $f_n$ for an inhomogeneous system
with a certain $n_{\rm peak}$, and find all observed spin dynamics can be fit by this formula
(see typical examples in Fig.~\ref{quenchQ}(a)),\vspace{-0.3pc}
\begin{align}
\rho_0(t)= & \sum\limits_{n=2}^{n_{\rm peak}}{A_{n}\exp(-t/\tau_n) \sin{[2\pi f_{n} (t-t_0)]}} \notag\\
&+\Delta\rho_{0}\exp(-t/\tau_0)+\frac{1}{3}~. \label{fitting}
\end{align}
Here the first term combines individual Rabi-type oscillations at all possible $n$ with $1/\tau_n$ being the damp rate for
oscillation amplitudes and $t_0$ marking the beginning of oscillations, while the second term describes an overall decay
of spin oscillations at a decay rate of $1/\tau_0$. This decay may be mainly due to unavoidable lattice-induced heatings.
The third term of Eq.~\eqref{fitting} is based on Refs.~\cite{Zhao2dLattice,Thermal} and indicates the three spin
components equally distribute in equilibrium states when $t_{\rm hold}\rightarrow \infty$.

To better illustrate the spin-mixing dynamics, we conduct fast Fourier transformations (FFT) onto all observed time
evolutions. Two typical FFT spectra extracted from the same data set over different time durations are shown in
Fig.~\ref{FFT}(a), where the vertical lines mark the five $f_n$ predicted by MFT. Each of these two FFT spectra has five
distinguished peaks agreeing well with the MFT predictions, i.e., all spin components in the three even Mott lobes
oscillate at lower frequencies while particles in the two odd Mott lobes display higher spin oscillation frequencies when
$q/U_2 < 1.55$. Atom number distributions in the spinor gases can also be revealed from the corresponding FFT spectrum
over a given time duration, as explained in Figs.~\ref{FFT}(b) and \ref{FFT}(c). A comparison between these two figures
clearly demonstrates that number distributions $\chi_n$ in our system quickly change with time $t_{\rm hold}$ and the
$n=2$ Mott lobe becomes more dominating after atoms are held in deep lattices for a longer time. This implies atoms in the
$n=2$ Mott lobe decay more slowly, which may be owing to a lack of three-body inelastic collisions in this lobe.
Figure~\ref{FFT}(b) shows another notable result: each experimental $\chi_n$ extracted from the FFT spectrum over a short
time duration (i.e., $t_{\rm hold}=40$~ms) coincides with the theoretical $\chi_n$ derived from Eq.~\eqref{MF} and the
Thomas-Fermi approximation for Mott-insulator shells at $n_{\rm peak}=6$. Atoms in initial states distribute into these
predicted Mott shells during the Quench-Q sequences, because the initial states are the ground states of the MI phase. Our
data thus experimentally confirm that the spin-mixing dynamics and their corresponding FFT spectra over a short $t_{\rm
hold}$ can efficiently probe the initial Fock-state distributions after a sufficiently fast quench.

Similar non-equilibrium dynamics composed of various spin-mixing oscillations are also detected in time evolutions of
spinor gases after Quench-L sequences under a wide range of magnetic fields, as shown in Fig.~\ref{U2}. To our knowledge,
this may be the first experimental observation of such complicated spin-mixing dynamics, although its theoretical model
has been studied by Ref.~\cite{spinorquench}. Our observations indicate the spin-mixing dynamics weakly depend on $t_{\rm
ramp}$~\cite{relax}. Typical examples can be seen in Fig.~\ref{U2}(a), where the data sets collected at distinct $t_{\rm
ramp}$ display similar dynamics with almost identical oscillation frequencies and slightly different oscillation
amplitudes. This may be due to the fact that $t_{\rm ramp}$ in a Quench-Q sequence is carefully chosen for limiting all
spin components to oscillate between the ground states and the first excited states.

The spin oscillations observed after Quench-L sequences can also be well fit by Eq.~\eqref{fitting} (see
Fig.~\ref{U2}(a)). We can thus extract the spin-dependent interaction $U_2$ from these fitting curves, because the
oscillation frequencies $f_n$ are decided by $U_2$ when $n\geq 2$ at a fixed $q$. Figures~\ref{U2}(b) and \ref{U2}(c) show
20 experimental values of $U_2$ extracted from our data taken under very different conditions. By applying linear fits to
these data points, we find a precise value for two key parameters that determine the spinor physics, i.e., $ U_2/U_0\simeq
0.035 (3)$ and $a_2/a_0\simeq 1.115 (10)$ for $^{23}$Na atoms. Here $a_2$ and $a_0$ are s-wave scattering lengths, and
$a_2/a_0=(U_2+U_0)/(U_0-2U_2)$ based on Ref.~\cite{Ho1998,Machida1998}. Many published values of $U_2/U_0$ were derived from the
scattering lengths~\cite{spinorquench,photoassociation,Na2Scattering,FeshbachSpec,EiteSpec,U2U01,U2U02,U2U03}. For example,
Refs.~\cite{photoassociation,Na2Scattering} respectively found scattering lengths that would lead to $U_2/U_0=0.032(14)$
and $0.035(11)$. In addition, experimental measurements of the scattering lengths through Feshbach spectroscopy could
yield $U_2/U_0=0.037(6)$~\cite{EiteSpec} and $0.036(3)$~\cite{FeshbachSpec}. Therefore, the observed spin dynamics can
conveniently measure spin-dependent interactions and $U_2/U_0$ with a good resolution.

We also notice one puzzling difference between the non-equilibrium dynamics initiated by a Quench-L sequence and those via
a Quench-Q sequence: atoms appear to oscillate with a larger amplitude despite having the same frequencies after the
Quench-Q sequence, even if spinor gases are prepared into the same final $u_L$ and $q$ by these two quench sequences. This
amplitude difference may be attributed to the inevitable dephasing and energy dissipations induced by a number of
tunnelling processes. Note that atoms are fully localized in individual lattice sites with negligible tunnellings during
Quench-Q sequences. In contrast, spinor gases cross SF-MI phase transitions during a Quench-L sequence, tunnellings among
adjacent sites thus cannot be ignored during a certain part of this sequence. Other possible reasons for the different
oscillation amplitudes may include significant heatings induced by first-order SF-MI phase transitions at a small $q$
during Quench-L sequences~\cite{JiangLattice}, different atom number distributions introduced by the quench
sequences~\cite{quenchL}, and non-adiabatic lattice ramps in Quench-L sequences.

\begin{figure}[t]
\includegraphics[width=85mm]{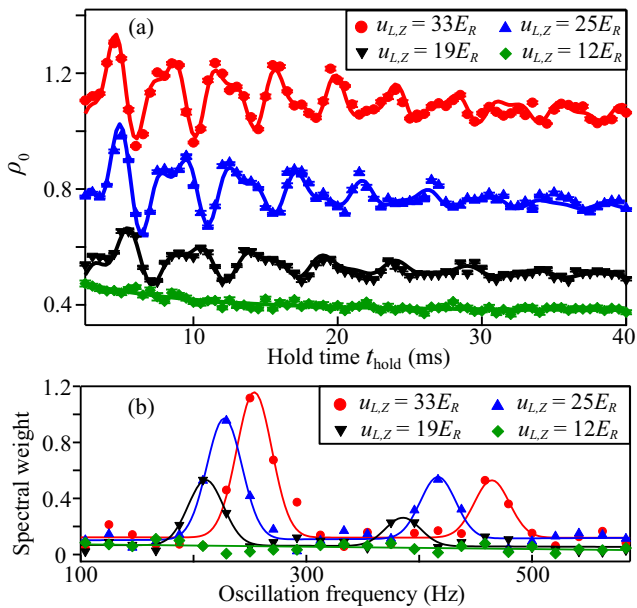}
\caption{(a) Observed spin dynamics after Quench-Q sequences to $q/h=30$~Hz at various $u_{L,z}$ while
$u_{L,x}=u_{L,y}=33E_{R}$ (see text). Results obtained at $u_{L,z}=33E_{R}$, $25E_{R}$, and $19E_{R}$ are respectively
shifted up by 0.55, 0.25, and 0.06 for visual clarity. Lines are fits based on Eq.~\eqref{fitting}. (b) FFT spectra of the
dynamics shown in Panel~(a). Lines are two-Gaussian fits. } \label{uL}
\end{figure}

To understand how tunnellings affect the spin-mixing dynamics, we monitor spin oscillations after varying the tunnelling
energy $J$ in a well-controlled way~\cite{KSpinDynamics}. We first prepare a non-equilibrium initial state with a Quench-Q
sequence to $q/h=30$~Hz in a very deep cubic lattice of $u_{L,x}=u_{L,y}=u_{L,z}=33E_R$ with $J\simeq 0$; and then
suddenly increase $J$ to a desired value by properly reducing only one lattice depth $u_{L,z}$. Here $u_{L,x}$, $u_{L,y}$,
and $u_{L,z}$ are depths of the three lattice beams along orthogonal directions, respectively. Interesting results shown
in Fig.~\ref{uL} are collected at four signature $u_{L,z}$, gradually spanning from the few-body dynamics for spinor gases
tightly localized in deep lattices at $u_{L,z}=33E_R$ with $J\simeq 0$, to the many-body dynamics for atoms loosely
confined in shallow lattices with $J\gg 0$ at $u_{L,z}=12E_R$. Amplitudes of spin-mixing oscillations appear to quickly
decrease as $u_{L,z}$ is reduced, and completely vanish when $u_{L,z}<14E_R$. We may understand these observations from
two simple illustrations. In one scenario, two atoms oscillate at the frequency $f_2$ in an $n=2$ lattice site. The spin
oscillation disappears as one of the two atoms tunnels out of the site. In another scenario, $n>2$ atoms oscillate in a
lattice site at the frequency $f_n$. After one atom hopping out of this site, spin oscillations occurring in this site and
the adjacent site that accepts the atom should be changed. Many of such tunnelling events could significantly reduce
oscillation amplitudes of the observed spin-mixing dynamics. As $J$ increases with the reduction of $u_{L,z}$, the damping
is enhanced and eventually stops the spin oscillations. As a numerical example, the predicted time scale corresponding to
$J$ along the $z$-direction is around 3~ms at $u_{L,z}=12E_{R}$, which is comparable to the damp rate of the observed
oscillations (see Fig.~\ref{uL}(a)). These results justify our use of deep lattices and subsequent neglecting of $J$ in
Eq.~\eqref{MF}.

Figure~\ref{uL}(b) show the FFT spectra extracted from the non-equilibrium dynamics observed at the four $u_{L,z}$. Each
of these FFT spectra has only two distinguished peaks rather than the predicted five peaks, i.e., the wide peaks at around
250~Hz correspond to the oscillations of even $n$ atoms and the wide peaks at around 450~Hz to the oscillations of odd $n$
atoms. One possible reason for this discrepancy is $t_{\rm hold}$ needs to be much longer (greater than 160~ms for all
even $n$) to reduce the aliasing effect of the spectrum analysis, but $t_{\rm hold}$ in our system is limited by lattice
heatings and atom losses. The FFT spectra in Fig.~\ref{uL}(b), however, clearly show that a larger $ u_{L,z} $ leads to
spin oscillations of higher frequencies. This can be interpreted by the fact that the oscillation frequency $f_n$ is
determined by $ U_2 $ and thus also by the effective lattice depth $u_L=\sqrt[3]{u_{L,x}u_{L,y}u_{L,z}}$. Our calculations
confirm that the effective $U_2$ gives oscillation frequencies that fall into those broad peaks seen in Fig.~\ref{uL}(b).

In conclusion, we have presented the first experimental study on few-body spin dynamics and transitions between the
well-studied two-body and many-body dynamics in antiferromagnetic spinor BECs. Intricate dynamics consisting of
spin-mixing oscillations at multiple frequencies have been observed in time evolutions of the spinor condensate localized
in deep lattices after two different quantum quench sequences. We have confirmed these observed spectra of spin-mixing
dynamics can reveal atom number distributions of an inhomogeneous system and also enable precise measurements of two key
parameters. The lattice quench method is applicable to other spinor systems, although antiferromagnetic spinor BECs may display larger spin oscillation amplitudes than ferromagnetic spinor
gases~\cite{spinorquench}.

\begin{acknowledgments}
We thank Eite Tiesinga for insightful discussions. We also thank the National Science Foundation, the Noble Foundation,
and the Oklahoma Center for the Advancement of Science and Technology for financial support.
\end{acknowledgments}


\begin{thebibliography}{100}
\bibitem{procedure} We prepare the LP state by eliminating $|F=1, m_F=\pm1\rangle$ atoms with resonant microwave and laser
pulses~\cite{ZhaoSinglet}. The setup and calibration of cubic lattices are identical to those illustrated in our previous
work~\cite{JiangLattice}.
\bibitem{ramp} First, $t_{\rm ramp}$ should be long enough to satisfy the interband adiabaticity requirement
$d u_L/dt\ll 32\pi E_R^2/h$~\cite{ZhaoSinglet}. Second, $ t_{\rm ramp} $ must be short enough to initiate spin-mixing
oscillations, while sufficiently long to limit the oscillations between the ground states and the first excited states.
\bibitem{FFT} Our calculations based on Eq.~\eqref{MF} indicate spin oscillation ampltitudes $D_n$ for an inhomogeneous system
with $n_{\rm peak}=6$ at $q/h=85$~Hz are $D_1=0$, $D_2=0.413$, $D_3=0.229$, $D_4=0.269$, $D_5=0.215$, and $D_6=0.199$.
Because no spin oscillations occur when $n=1$, $\chi_n$ shown in Fig.~\ref{FFT} reflect the normlized number distributions
after the $n=1$ Mott lobe is excluded.
\bibitem{relax} When $t_{\rm ramp}< 1$~ms, we find atoms in our system need additional 1~ms to completely lose their phase coherence. The similar phenomenon has
also been reported in rubidium systems~\cite{NumState}.
\end{thebibliography}
\end{document}